\documentclass[12pt]{article}
\usepackage{fullpage,
,epsfig,psfrag,graphics,amsbsy,amssymb,amsmath
}
\usepackage{caption}
\newcommand{\beq}{\begin{equation}}
\newcommand{\eeq}{\end{equation}}
\newcommand{\bea}{\begin{eqnarray}}
\newcommand{\eea}{\end{eqnarray}}
\newcommand{\bfs}{\boldsymbol}

\newcommand{\Tr}{{\rm Tr}}
\newcommand{\be}{\begin{equation}}
\newcommand{\ee}{\end{equation}}
\newcommand{\bq}{\begin{eqnarray}}
\newcommand{\eq}{\end{eqnarray}}
\newcommand{\ket}[1]{|#1\rangle}
\newcommand{\bra}[1]{\langle#1|}

\def\math{\mathsurround=0pt }
\def\leftrightarrowfill{$\math \mathord\leftarrow \mkern-6mu
 \cleaders\hbox{$\mkern-2mu \mathord- \mkern-2mu$}\hfill
 \mkern-6mu \mathord\rightarrow$}
\def\overleftrightarrow#1{\vbox{\ialign{##\crcr
     \leftrightarrowfill\crcr\noalign{\kern-1pt\nointerlineskip}
     $\hfil\displaystyle{#1}\hfil$\crcr}}}

\newcommand{\VEV}[1]{\langle#1\rangle}

\let\l=\lambda

 \def\bd{\begin{document}} \def\ed{\end{document}}
\def\ds{\documentstyle} \let\fr=\frac \let\bl=\bigl \let\br=\bigr
\let\Br=\Bigr \let\Bl=\Bigl
\let\bm=\bibitem
\let\na=\nabla
\let\pa=\partial \let\ov=\overline
\def\ft#1#2{{\textstyle{{\scriptstyle #1}\over {\scriptstyle #2}}}}
\def\fft#1#2{{#1 \over #2}}
\def\vp{\varphi}
\def\sst#1{{\scriptscriptstyle #1}}
\def\oneone{\rlap 1\mkern4mu{\rm l}}
\def\td{\tilde}
\def\wtd{\widetilde}
\def\dalemb#1#2{{\vbox{\hrule height .#2pt
        \hbox{\vrule width.#2pt height#1pt \kern#1pt
                \vrule width.#2pt}
        \hrule height.#2pt}}}
\def\square{\mathord{\dalemb{6.8}{7}\hbox{\hskip1pt}}}
\def\wtd{\widetilde}
\def\R{\rlap{\rm I}\mkern3mu{\rm R}}
\def\im{{\rm i}}
\def\tilg{\tilde{g}}
\def\tilF{\tilde{F}}
\def\tilA{\tilde{A}}
\def\varf{\varphi}
\def\tilf{\tilde{\phi}}
\def\tilh{\tilde{h}}
\def\rme{{\rm e}}
\def\ep{\epsilon}
\def\0{{(0)}}
\def\9{{(9)}}
\def\8{{(8)}}
\def\7{{(7)}}
\def\6{{(6)}}
\def\5{{(5)}}
\def\4{{(4)}}
\def\3{{(3)}}
\def\2{{(2)}}
\def\1{{(1)}}
\newcommand{\trace}{{\rm Tr}}
\newcommand{\ub}{\overline{U}}
\newcommand{\vb}{\overline{V}}
\newcommand{\uh}{\widehat{U}}
\newcommand{\vh}{\widehat{V}}
\newcommand{\ubh}{\overline{\widehat{U}}}
\newcommand{\vbh}{\overline{\widehat{V}}}
\newcommand{\lb}{\bar{\l}}
\newcommand{\Fb}{\overline{F}}
\newcommand{\Fh}{\widehat{F}}
\newcommand{\Fbh}{\overline{\widehat{F}}}
\newcommand{\Ab}{\overline{A}}
\newcommand{\Ah}{\widehat{A}}
\newcommand{\Abh}{\overline{\widehat{A}}}
\newcommand{\Gb}{\overline{G}}
\newcommand{\Gh}{\widehat{G}}
\newcommand{\Gbh}{\overline{\widehat{G}}}
\newcommand{\Pb}{\overline{P}}
\newcommand{\Ph}{\widehat{P}}
\newcommand{\Pbh}{\overline{\widehat{P}}}
\newcommand{\Qb}{\overline{Q}}
\newcommand{\Qh}{\widehat{Q}}
\newcommand{\Qbh}{\overline{\widehat{Q}}}
\newcommand{\Bb}{\overline{B}}
\newcommand{\Bh}{\widehat{B}}
\newcommand{\Bbh}{\overline{\widehat{B}}}
\newcommand{\fhns}{\hat{F}^{\rm (NS)}}
\newcommand{\fhrr}{\hat{F}^{\rm (RR)}}
\newcommand{\ahns}{\hat{A}^{\rm (NS)}}
\newcommand{\ahrr}{\hat{A}^{\rm (RR)}}
\newcommand{\hhrr}{\hat{H}^{\rm (RR)}}
\newcommand{\hchi}{\hat{\chi}}
\newcommand{\hphi}{\hat{\phi}}
\newcommand{\htau}{\hat{\tau}}
\newcommand{\cG}{{\cal G}}
\newcommand{\cGb}{\overline{{\cal G}}}
\newcommand{\cH}{{\cal H}}
\newcommand{\cP}{{\cal P}}
\newcommand{\cPb}{\overline{{\cal P}}}
\newcommand{\cQ}{{\cal Q}}
\newcommand{\cQb}{\overline{{\cal Q}}}
\newcommand{\cM}{{\cal M}}
\newcommand{\cN}{{\cal N}}
\newcommand{\cO}{{\cal O}}
\newcommand{\cD}{{\cal D}}
\newcommand{\cL}{{\cal L}}

\newcommand{\vpp}{\mbox{$\langle{\scriptstyle++}\rangle$}}
\newcommand{\vmp}{\mbox{$\langle{\scriptstyle-+}\rangle$}}
\newcommand{\vppp}{\mbox{$\langle{\scriptstyle+++}\rangle$}}
\newcommand{\vmpp}{\mbox{$\langle{\scriptstyle-++}\rangle$}}
\newcommand{\vpmp}{\mbox{$\langle{\scriptstyle+-+}\rangle$}}

\begin{document}
\setlength{\captionmargin}{36pt}
\begin{titlepage}
\begin{flushright}
\phantom{UFIFT-HEP}
\end{flushright}

\vskip 3cm
\begin{center}
\begin{large}
{\bf Stable String Bit Models}
\end{large}

\vskip 2cm
{\large
Songge Sun\footnote{E-mail  address: {\tt uranussg@phys.ufl.edu}} 
and Charles B. Thorn\footnote{E-mail  address: {\tt thorn@phys.ufl.edu}}
}
\vskip0.20cm
{\it Institute for Fundamental Theory\\
Department of Physics, University of Florida,
Gainesville FL 32611}


\vskip 1.0cm
\end{center}

\begin{abstract}
\noindent In string bit models, the superstring emerges as a very
long chain of ``bits'', in which $s$ fermionic 
degrees of freedom contribute positively to the ground state energy in
a way to exactly cancel the destabilizing negative contributions of
$d=s$ bosonic degrees of freedom. We propose that the physics of
string formation be studied nonperturbatively
in the class of string bit models in which $s>d$, so that
a long chain is stable, in contrast to the marginally stable 
($s=d=8$) superstring chain.   
We focus on the simplest of these models with $s=1$ and $d=0$, in which 
the string bits  live in zero space dimensions.
The string bit creation operators are $N\times N$
matrices. We choose a Hamiltonian such that the large $N$ limit produces
string moving in one space dimension, with excitations corresponding
to one Grassmann lightcone worldsheet field ($s=1$) and no bosonic 
worldsheet field ($d=0$).
We study this model at finite $N$ to assess the role of
the large $N$ limit in the  emergence of the spatial
dimension. Our results suggest
that string-like states with large bit number $M$ 
may not exist for $N\leq (M-1)/2$. 
If this is correct,  one can have finite chains of string bits, 
but not continuous string, at finite $N$. Only for extremely large $N$ 
can such chains behave approximately like continuous string,
in which case there will also be the (approximate) emergence of a new 
spatial dimension.
In string bit models designed to produce critical superstring
at $N=\infty$, we can then expect only approximate
Lorentz invariance at finite $N$, with violations of order $1/N^2$. 
\end{abstract}
\vfill
\end{titlepage}
\section{Introduction}
String bit models \cite{thornsakh} provide one approach 
to a fundamental formulation of string theory. For another 
approach see \cite{thooftfound}.
These models are motivated by interpreting
the lightcone Hamiltonian for a single string
\cite{goddardgrt}
\bea
P^-&=& \frac{1}{2}\int_0^{P^+} d\sigma \left[{\bfs p}^2+T_0^2{\bfs x}^{\prime2}
\right]
\eea
as the large $M$ limit of the Hamiltonian for a harmonic chain of $M
=P^+/m$ string bits\footnote{The compatibility of such a discretization with string interactions
is supported by the success of a similar discretization of 
Mandelstam's interacting
lightcone worldsheet path integrals \cite{mandelstamlc} carried out 
in \cite{gilest}.}, where $m$ is the fundamental unit of $P^+$
\bea
H&=&\frac{1}{2m}\sum_{k=1}^M\left[{\bfs p}_n^2+T_0^2({\bfs x}_{n+1}
-{\bfs x}_n)^2\right].
\eea 
The idea is to take string bits as the fundamental
degrees of freedom of string theory. In this interpretation Lorentz
invariance is not built in {\it a priori}. Moreover, string bits
move about in the $d=D-2$ transverse space dimensions: the spatial
coordinate $x^-$ conjugate to $P^+$ is missing. It will be regained
for strings consisting of a very large number of bits, provided that
the excitation spectrum of $H$ scales like $1/M=m/P^+$
in the limit $M\to\infty$. This is one of the earliest
implementations of 't Hooft's holography hypothesis \cite{thoofthologram}. 
The normal mode frequencies of the closed string bit chain are
$\omega_n=(2T_0/m)\sin(n\pi/M)$, with $n=0,\ldots,(M-1)$.
Thus all modes with finite
$n$ or finite $M-n$ as $M\to\infty$ have the desired scaling behavior.

\begin{figure}[ht]
\begin{center}
\includegraphics[width=4in]{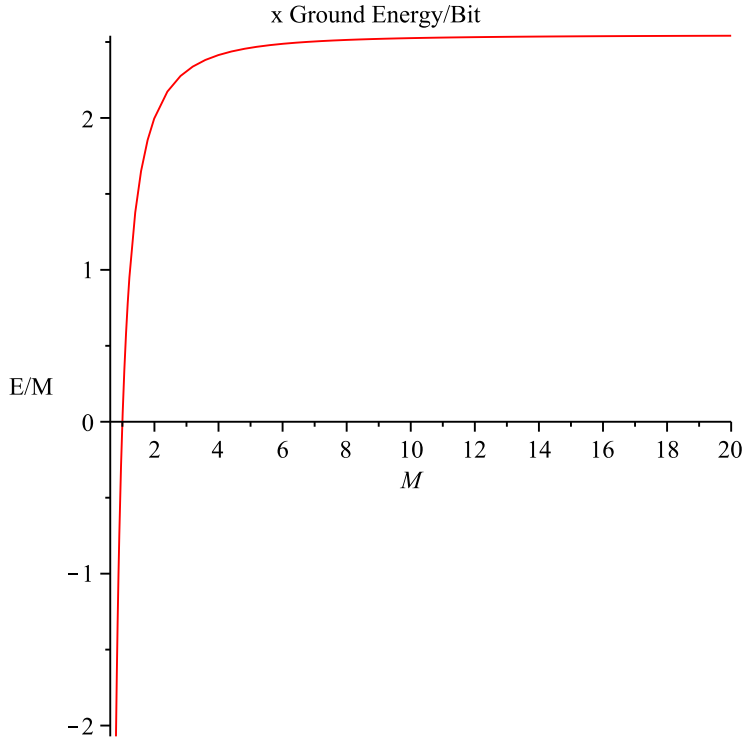}
\caption{Energy per bit number versus bit number}
\label{enperbitx}
\end{center}
\end{figure}

But the ground state energy of a bosonic closed string bit chain is
\bea
E_G&=&\frac{1}{2}\sum_n\omega_n
=\frac{dT_0}{m}\sum_{n=1}^{M-1}\sin\frac{n\pi}{M}
=\frac{dT_0}{m}\cot\frac{\pi}{2M}=\frac{2dT_0M}{m\pi}-\frac{\pi dT_0}{6Mm}
+{\mathcal O}(M^{-3})
\eea
where we have taken $M\to\infty$ to regain the continuous string.
Since the string interactions conserve bit number, the first term can
be dropped \cite{gilest}, and the second term can be identified
as $-\pi dT_0/(6P^+)$. For $d=24$ we obtain the ground state mass
squared of the bosonic string $-8\pi T_0$. This
result is a version of the zero point energy calculation
of Brink and Nielsen \cite{brinknielsen}.
Since the ground state is a tachyon, it signals the instability 
of bosonic string theory. 
Moreover, this instability is present at all finite $M$,
which is immediately apparent from the monotonicity of the graph of
ground energy per bit number $E_G(M)/M$ 
as a function of bit number shown in 
Fig.~\ref{enperbitx}.
For a fixed total number of bits, this graph implies that the lowest 
energy configuration is for the bits to arrange themselves into
closed chains of least possible bit number. In other words, 
if the interaction between closed chains is weak, a closed chain 
of string bits is unstable. Or when a number of bits are brought together they
remain individual bits, never joining to form string.
In short, string cannot be expected to emerge from
the bosonic string bit model.

This is hardly an unexpected conclusion, since bosonic string is widely 
believed to be unstable. This inevitable instability is one
motivation for introducing supersymmetry \cite{gso}.
One imagines that in addition to the vibrational
excitations discussed above, there are also fermionic degrees of
freedom \cite{bardakcihalpern,greenschwarz}, which are
properly thought of as ``statistics waves'', 
and which contribute to the ground energy with the opposite sign, 
cancelling the negative $1/M$ term and leading to massless ground states. 
Bergman and one of us invented a superstring bit model which accomplishes this
cancellation at finite $M$ \cite{bergmantsubit}. The
ground state energy of a chain of superstring bits
is strictly zero for all finite $M$.

Employing
't Hooft's large $N$ limit \cite{thooftlargen} in its 
Fock space formulation \cite{thornfock}, as
described in \cite{thornsakh} for the bosonic string, we introduced 
in \cite{bergmantsubit} a 
superstring bit annihilation operator
\bea
(\phi_{[a_1\cdots a_n]})_\alpha^{~\beta}({\bfs x}),\qquad n=0,\dots, s
\eea
where each $a_i$ is a spinor index running over $s$ values, and
$\alpha,\beta=1,\ldots N$ are color indices for the adjoint representation 
of the color group $U(N)$. Poincar\'e supersymmetry dictates that $s=d=8$
for the superstring. The $\phi$'s are bosonic if $n$ is even and
fermionic if $n$ is odd. The square brackets in the subscript remind us that
the enclosed indices are completely antisymmetric. Thus at each
transverse space point ${\bfs x}$ there are 256 degrees of freedom,
128 each of bosonic and fermionic type. We then
constructed a Hamiltonian operator which reproduced, in the
't Hooft limit, the mass
spectrum of the free superstring when the
bit number becomes very large\footnote{This conclusion rigorously
follows only when $N\to\infty$ {\it before} $M$ becomes large.}. 
The problem
of devising the correct superstring interactions, which should
be consistent with Lorentz invariance, was only partially
resolved in \cite{bergmantsubit}. But temporarily setting aside
Lorentz invariance, we stress that our model did induce a (Lorentz
non-invariant) interacting superstring theory along with a new 
dimension of space.

The ground state energy of a closed chain
in this superstring bit model is exactly
zero for all $M$, but only because of the cancellation between
phonons and statistics waves \cite{thornsubstructure}. 
This cancellation does not occur
if $s\neq d$. In this more general context we can identify
a stable regime ($s>d$) and an unstable regime ($s<d$).
The supersymmetric case is at the boundary between these two regimes.
Here we are proposing that, since superstring theory emerges from
a very special, marginally stable, string bit model, its underlying physics 
is better understood in terms of the more general class of stable string bit 
models. Indeed, we can regard holding $d=D-2<s$ as 
a physical infrared cutoff, in the spirit of dimensional regularization.

\begin{figure}[ht]
\begin{center}
\includegraphics[width=4in]{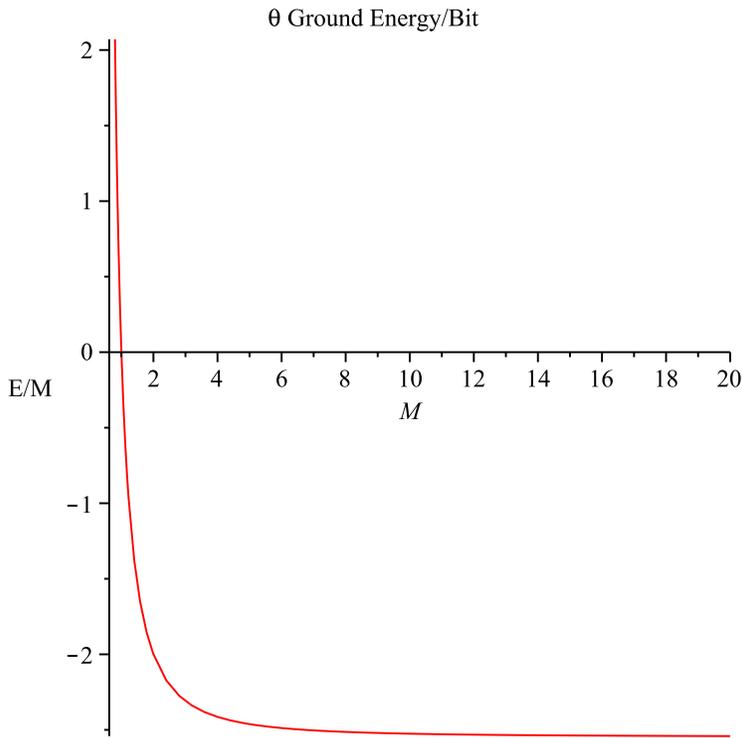}
\caption{Energy per bit number versus bit number}
\label{enperbittheta}
\end{center}
\end{figure}

In these stable models the energy per bit curve gets flipped and
turns into Fig.\ref{enperbittheta}.
This figure shows that with fixed $M$ the lowest energy state
is a single string, which becomes a string moving in 1
space dimension when $M\to\infty$. If two closed strings are present the
lowest energy configuration shares $M$ equally between the
two strings.
Thus models in the stable regime provide a sound foundation for the
emergence of string, albeit without Lorentz invariance.
In the rest of this article we analyze the physics of the simplest of
these stable models, with $s=1$ and $d=D-2=0$. Holography
is explicitly realized in the 't Hooft limit $N\to\infty$.
We will be particularly interested in the physics of these models at finite
$N$, and whether finite $N$ can still support 
chains of arbitrarily large bit
number. Indeed, we find indications, from our study of
low dimension toy string bit models, that a stable string of $M$
bits requires {\it at least} $N>(M-1)/2$. If this is so, there will be 
violations of the Lorentz invariant dispersion law $2P^+P^--{\bfs p}^2=m^2$ 
at finite $N$, 
because $P^+$ would necessarily remain discrete at finite $N$.
These violations would be present even if the degrees of freedom and 
interactions are tuned to satisfy Lorentz invariance order by order
in the $1/N$ expansion, which these considerations suggest
has zero radius of convergence. One should then be able to translate
limits on the accuracy of tests of Lorentz invariance to a
lower bound on $N$, or, if a successful string bit model
of all physics could be devised, to an upper bound on
Newton's gravitational constant, which is of order $1/N^2$ in these
models.

The superstring bit model
in zero space dimensions with $s=1$, studied in the
rest of this paper, is a far cry from
a successful model of all physics! Nonetheless it is rich enough to
test the soundness of the physical ideas we are advocating. 
In Section 2 we present
the details of the model. Among many possible choices for a Hamiltonian
built of single trace operators we single out the one proposed
in \cite{bergmantsubit} and discuss its $N\to\infty$ limit. 
In Section 3, we apply the variational principle to obtain
an upper bound on the ground state energy. This bound can be proven only when
$N>(M-1)/2$.
Then in Section 4 we explore the model at finite $N$ by studying
states with relatively low bit number $M$. We find all energy eigenstates 
in the 2 bit case $M=2$ and all color singlet and color adjoint states when 
$M=3$. We also describe some
results for $M=4,5$, for which the detailed analysis will appear in a 
separate article. We close the paper in Section 5 with a discussion
of the issues involved with finding string bit models that produce
string in higher dimensions. In the interests of keeping the paper
self-contained we include an appendix which calculates the 
exact energy eigenstates at $N=\infty$ for the model studied here. This
is of course a special case of results already obtained in \cite{bergmantsubit}
in a general context. However it includes two new
aspects: (1) the explicit analysis of the
consequences of the cyclic symmetry of closed chains of string bits,
and (2) the calculation of the $N=\infty$
energy eigenvalues of an open (color adjoint) chain, which demonstrates
color confinement in this toy string bit model. The energy gap between
adjoint and singlet sectors is of order $M$ times the energy scale
set by the continuum limit. We also include a second appendix which
sketches a systematic method to do variational calculations in these
models. 
\section{The $s=1$, $d=0$ String Bit Model}
The $s=1$, $d=0$ superstring bit model contains $N^2$ 
bosonic ($a_\alpha^\beta$) and $N^2$ fermionic 
($b_\alpha^\beta$) string bits. We define conjugate operators
${\bar a}_\alpha^\beta\equiv (a_\beta^\alpha)^\dagger$ and
${\bar b}_\alpha^\beta\equiv (b_\beta^\alpha)^\dagger$, and these
operators satisfy the (anti-)commutation relations:
\bea
{}[a_\alpha^\beta,{\bar a}_\gamma^\delta]&=&
\delta_\alpha^\delta\delta_\gamma^\beta,
\qquad\{b_\alpha^\beta,{\bar b}_\gamma^\delta\}=
\delta_\alpha^\delta\delta_\gamma^\beta
\eea
all others vanishing. Then we choose a Hamiltonian that is a linear
combination of the single trace operators 
\bea
\Tr\ {\bar a}^2a^2,\quad\Tr\ {\bar b}^2b^2,\quad\Tr\ {\bar b}^2a^2,
\quad\Tr\ {\bar a}^2b^2,\quad\Tr\ {\bar a}{\bar b}ba,\quad\Tr\ {\bar a}{\bar b}ab,
\quad\Tr\ {\bar b}{\bar a}ba,\quad\Tr\ {\bar b}{\bar a}ab
\eea
with coefficients scaling as $1/N$.
All of these structures conserve bit number and share the feature that
the two annihilation operators are consecutive in the trace as are
the two creation operators. This feature is necessary for
the term to survive
the limit $N\to\infty$. Terms without this feature such as
$(1/N)\Tr:{\bar a}a{\bar a}a:$ are suppressed at large $N$, so they can
also be included, if desired, without affecting the large $N$ limit.
We don't include them here for simplicity only, but in some models
they may be necessary for stability reasons.

For definiteness we choose $H$ so that in
the large $N$ limit the dynamics reduces to the superstring bit
model invented in \cite{bergmantsubit}. In the large $N$ limit,
$H$ maps single trace states to single trace states. To describe
these states in this toy model, it is useful to define a super
creation operator
\bea
\psi(\theta)&=&{\bar a}+{\bar b}\theta,\qquad {\bar b}=-\frac{d}{d\theta}\psi,
\quad {\bar a}=\left(1-\theta\frac{d}{d\theta}\right)\psi
\eea
where $\theta$ is a Grassmann anti-commuting number. Then a basis
of single trace states can be taken to be
\bea
\ket{\theta_1\theta_2\cdots\theta_M}&=&
\Tr\ [\psi(\theta_1)\psi(\theta_2)\cdots\psi(\theta_M)]\ket{0}
\eea
Then it is straightforward to apply each of the candidate terms in
the Hamiltonian to
such a basis state to get 
\bea
\frac{1}{N}\Tr\ {\bar a}^2a^2\ket{\theta_1\theta_2\cdots\theta_M}&=&
\sum_{k=1}^M \left(1-\theta_k\frac{d}{d\theta_k}\right)\left(1-\theta_{k+1}
\frac{d}{d\theta_{k+1}}\right)\ket{\theta_1\theta_2\cdots\theta_M}
+{\cal O}(N^{-1})
\label{dict1}\\
\frac{1}{N}\Tr\ {\bar a}{\bar b}ba\ket{\theta_1\theta_2\cdots\theta_M}&=&
\sum_{k=1}^M\left(1- \theta_k\frac{d}{d\theta_k}\right)\theta_{k+1}
\frac{d}{d\theta_{k+1}}\ket{\theta_1\theta_2\cdots\theta_M}+{\cal O}(N^{-1})
\label{dict2}\\
\frac{1}{N}\Tr\ {\bar b}{\bar a}ab\ket{\theta_1\theta_2\cdots\theta_M}&=&
\sum_{k=1}^M \theta_k\frac{d}{d\theta_k}\left(1-\theta_{k+1}
\frac{d}{d\theta_{k+1}}\right)\ket{\theta_1\theta_2\cdots\theta_M}
+{\cal O}(N^{-1})\label{dict3}\\
\frac{1}{N}\Tr\ {\bar b}^2b^2\ket{\theta_1\theta_2\cdots\theta_M}&=&
\sum_{k=1}^M \theta_k\frac{d}{d\theta_k}\theta_{k+1}
\frac{d}{d\theta_{k+1}}\ket{\theta_1\theta_2\cdots\theta_M}+{\cal O}(N^{-1})
\label{dict4}
\eea\bea
\frac{1}{N}\Tr\ {\bar a}{\bar b}ab\ket{\theta_1\theta_2\cdots\theta_M}&=&
\sum_{k=1}^M \theta_k\frac{d}{d\theta_{k+1}}
\ket{\theta_1\theta_2\cdots\theta_M}+{\cal O}(N^{-1})\label{dict5}\\
\frac{1}{N}\Tr\ {\bar b}{\bar a}ba\ket{\theta_1\theta_2\cdots\theta_M}&=&
\sum_{k=1}^M \theta_{k+1}\frac{d}{d\theta_{k}}
\ket{\theta_1\theta_2\cdots\theta_M}+{\cal O}(N^{-1})\label{dict6}\\
\frac{1}{N}\Tr\ {\bar b}^2a^2\ket{\theta_1\theta_2\cdots\theta_M}&=&
\sum_{k=1}^M \frac{d}{d\theta_k}
\frac{d}{d\theta_{k+1}}\ket{\theta_1\theta_2\cdots\theta_M}+{\cal O}(N^{-1})
\label{dict7}\\
\frac{1}{N}\Tr\ {\bar a}^2b^2\ket{\theta_1\theta_2\cdots\theta_M}&=&
\sum_{k=1}^M \theta_{k+1}\theta_{k}\ket{\theta_1\theta_2\cdots\theta_M}
+{\cal O}(N^{-1})\label{dict8}
\eea
The structure of the $1/N$ terms not shown involves two traces rather than
a single trace. The formulas (\ref{dict6}), (\ref{dict7}), (\ref{dict3}), and
(\ref{dict4}) show a unique
correspondence between single trace terms and Grassmann variable 
operations. The remaining formulas possess some ambiguities in
the correspondence. Using (\ref{dict2}) we can remove the quartic
term from (\ref{dict1}), (\ref{dict5}) and (\ref{dict8}):
\bea
\frac{1}{N}\Tr\ \left[{\bar a}^2a^2-{\bar b}^2b^2\right]
\ket{\theta_1\theta_2\cdots\theta_M}&=&
\sum_{k=1}^M \left(1-2\theta_k\frac{d}{d\theta_k}\right)
\ket{\theta_1\theta_2\cdots\theta_M}
+{\cal O}(N^{-1})
\\
\frac{1}{N}\Tr\ \left[{\bar a}{\bar b}ba+{\bar b}^2b^2\right]
\ket{\theta_1\theta_2\cdots\theta_M}&=&
\sum_{k=1}^M\theta_{k}
\frac{d}{d\theta_{k}}\ket{\theta_1\theta_2\cdots\theta_M}+{\cal O}(N^{-1})
\\
\frac{1}{N}\Tr\ \left[{\bar b}{\bar a}ab+{\bar b}^2b^2\right]
\ket{\theta_1\theta_2\cdots\theta_M}&=&
\sum_{k=1}^M\theta_{k}
\frac{d}{d\theta_{k}}\ket{\theta_1\theta_2\cdots\theta_M}+{\cal O}(N^{-1})\eea
In the large $N$ limit we can make the {\it ansatz}
\bea
\ket{E}&=&\int d^M\theta\ket{\theta_1\theta_2\cdots\theta_M}
\Psi(\theta_1\theta_2\cdots\theta_M)
\eea
for the $N=\infty$ energy eigenstate. The function $\Psi$ is the
wave function for one of the discretized Grassmann variables of the
superstring. Now by construction the states $\ket{\theta_1\theta_2\cdots\theta_M}$ possess cyclic symmetry:
\bea
\ket{\theta_1\theta_2\cdots\theta_M}&=&\ket{\theta_M\theta_1\cdots\theta_{M-1}}
\eea
On the other hand the measure acquires a phase $(-)^{M-1}$ under a
one step cyclic transformation. It follows that the wave function
can, without loss of generality, be taken to satisfy
\bea
\Psi(\theta_1\theta_2\cdots\theta_M)&=&(-)^{M-1}
\Psi(\theta_M\theta_1\cdots\theta_{M-1})
\eea

Consulting  Eq. (3.17) of \cite{bergmantsubit}, we see that the first 
quantized Hamiltonian $h$ should be
\bea
h&=&\sum_{k=1}^M\left[-iS_kS_{k+1}+i{\tilde S}_k{\tilde S}_{k+1}-iS_k({\tilde S}_{k+1}
+{\tilde S}_{k-1}-2{\tilde S}_{k})\right]\nonumber\\
&=&\sum_{k=1}^M\left[-iS_kS_{k+1}+i{\tilde S}_k{\tilde S}_{k+1}
-iS_k{\tilde S}_{k+1}
-iS_{k+1}{\tilde S}_{k}+2iS_k{\tilde S}_{k})\right]
\eea
In these formulas it is understood that $S_{M+1}\equiv S_1$ and $S_0=S_M$.
Here the $S,{\tilde S}$ satisfy the Clifford algebras
\bea
\{S_k.S_l\}&=&2\delta_{kl},\qquad \{{\tilde S}_k.{\tilde S}_l\}=2\delta_{kl},
\qquad \{{S}_k,{\tilde S}_l\}=0
\eea
and have the representations
\bea
S_k&=&\theta_k+\frac{d}{d\theta_k},\qquad {\tilde S}_k=i\left(\theta_k
-\frac{d}{d\theta_k}\right)
\eea
in terms of Grassmann variables. Then we can also write
\bea
h&=&\sum_{k=1}^M\left[-2i\theta_k\theta_{k+1}
-2i\frac{d}{d\theta_k}\frac{d}{d\theta_{k+1}}
-2\theta_k\frac{d}{d\theta_{k+1}}-2\theta_{k+1}\frac{d}{d\theta_k}-2+4\theta_k\frac{d}{d\theta_k}\right]
\eea
Next we apply integration by parts to rewrite
\bea
\int d^M\theta\ket{\theta_1\theta_2\cdots\theta_M}
h\Psi(\theta_1\theta_2\cdots\theta_M)&=&
\int d^M\theta {\hat h}\ket{\theta_1\theta_2\cdots\theta_M}
\Psi(\theta_1\theta_2\cdots\theta_M)
\eea
where
\bea
{\hat h}&=&\sum_{k=1}^M\left[-2i\theta_k\theta_{k+1}
-2i\frac{d}{d\theta_k}\frac{d}{d\theta_{k+1}}
+2\theta_k\frac{d}{d\theta_{k+1}}
+2\theta_{k+1}\frac{d}{d\theta_k}+2-4\theta_k\frac{d}{d\theta_k}\right]
\eea
Now we compare the terms in ${\hat h}$ to the dictionary,
given in (\ref{dict1})--(\ref{dict8}), to infer the Fock space
Hamiltonian
\bea
H&=&\frac{2}{N}\Tr\left[i{\bar a}^2b^2
-i{\bar b}^2a^2
+{\bar a}{\bar b}ab
+{\bar b}{\bar a}ba+{\bar a}^2a^2-{\bar b}^2b^2\right]
\eea
Without affecting the $N\to\infty$ limit, we are free to add the terms
\bea
\Delta H&=&\frac{1}{N}\Tr\left[2\xi_1{\bar a}{\bar b}ba+2\xi_2{\bar b}{\bar a}ab
+(\xi_1+\xi_2)({\bar a}^2a^2+{\bar b}^2b^2-M)\right]
\eea
to $H$ because this combination of terms only contributes at order $N^{-1}$.
The number of bits in the state is given by
the bit number operator $M=\Tr({\bar a}a+{\bar b}b)$.

In the following we pick, for definiteness, $\xi_2=-\xi_1=1$, so
we will be using the Fock space Hamiltonian
\bea
H&=&\frac{2}{N}\Tr\left[({\bar a}^2 -i{\bar b}^2)a^2
-({\bar b}^2-i{\bar a}^2)b^2+({\bar a} {\bar b}
+{\bar b} {\bar a})ba+({\bar a} {\bar b}
-{\bar b} {\bar a})ab\right]
\label{bitham}
\eea
This Hamiltonian is supersymmetric in that it commutes with the
Grassmann odd operator $Q=\Tr({\bar a}be^{i\pi/4}+{\bar b}ae^{-i\pi/4})$.
This Hamiltonian will be our paradigm for the rest of the article.
A couple of comments are in order. As Appendix A explains, the spectrum
of $h$ is symmetric about $0$, {\it before} imposing the cyclic constraint,
which breaks this symmetry. This means that had we chosen $-H$
instead of $H$, the large $N$ limit would still be described by the
Grassmann variables of the superstring. However, in $-H$ the
negative coefficient of $\Tr{\bar a}^2a^2$ would cause a
dangerous instability at finite $N$ because $a$ is bosonic.
One can add more terms to make $-H$ stable, but we choose $+H$
as our paradigm, to keep the Hamiltonian as simple as possible. The negative
coefficient of $\Tr{\bar b}^2b^2$ in $H$ is not a problem because $b$
is fermionic and the exclusion principle stabilizes the
effects of this term.

\section{A Variational Argument}
In the previous section we have designed a string bit Hamiltonian
that reproduces free superstring dynamics at $N=\infty$. In this
section we seek finite $N$ information via the variational
principle. Inspection of the Hamiltonian (\ref{bitham}) suggests
that a low energy state should have a large number of fermionic excitations.
This encourages us to consider the trial state $\ket{\psi}
=\Tr{\bar b}^M\ket{0}$, which is nonzero only for $M$ odd, 
in the sector with bit number $M$, and 
evaluate
\bea
E(\psi)&=&\frac{\bra{\psi}H\ket{\psi}}{\VEV{\psi|\psi}}
\eea 
Since there are no ${\bar a}$ excitations, the numerator is easily
evaluated
\bea
\bra{\psi}H\ket{\psi}&=&-\frac{2}{N}\bra{0}\Tr b^M\Tr{\bar b}^2b^2
\Tr{\bar b}^M\ket{0}=-M\frac{2}{N}\bra{0}\Tr b^M{\bar b}^2b{\bar b}^{M-1}
\ket{0}
\nonumber\\
&=&-2M\bra{0}\Tr b^M\Tr{\bar b}^M\ket{0}-\frac{2M}{N}
\bra{0}\Tr b^M\Tr{\bar b}^2\sum_{k=1}^M\left[(-)^k\Tr{\bar b}^k\right]
{\bar b}^{M-k-2}\ket{0}\nonumber\\
&=&-2M\bra{0}\Tr b^M\Tr{\bar b}^M\ket{0}-\frac{2M}{N}
\bra{0}\Tr b^M\sum_{k=1}^M\left[(-)^k\Tr{\bar b}^k\right]
\Tr{\bar b}^{M-k}\ket{0}
\eea
Now $\Tr b^n=0$ for even $n$. Thus only the terms with $k$ odd contribute
to the sum on the right. But as already mentioned we must take $M$ odd
to get a nonzero trial, so $M-k$ is even implying $\Tr{\bar b}^{M-k}=0$.
Thus every term in the sum on the right vanishes and we have
\bea
\bra{\psi}H\ket{\psi}=-2M\bra{0}\Tr b^M\Tr{\bar b}^M\ket{0}=-2M\VEV{\psi|\psi}
\eea
So provided that $\VEV{\psi|\psi}\neq0$, we get the variational
estimate $E(\psi)=-2M$, so $E_G\leq -2M$.

It remains to calculate the norm of the trial state
\bea
\VEV{\psi|\psi}&=&\bra{0}\Tr b^M\Tr{\bar b}^M\ket{0}
=M\bra{0}\Tr b^{M-1}{\bar b}^{M-1}\ket{0}
\eea
To evaluate the right side we first derive a recursion formula:
\bea
\bra{0}\Tr b^{M-1}{\bar b}^{M-1}\ket{0}
&=&\bra{0}\Tr b^{M-2}\left[N{\bar b}^{M-2}+\sum_{k=1}^{M-2}
(-)^k\left[\Tr{\bar b}^k\right]{\bar b}^{M-k-2}\right]\ket{0}\nonumber\\
&=&\left(N^2-\sum_{k=odd}^{M-2}k\right)\bra{0}\Tr b^{M-3}{\bar b}^{M-3}\ket{0}
\nonumber\\
&&\bra{0}\Tr b^{M-3}\left[N\sum_{k=1}^{M-3}
(-)^k\left[\Tr{\bar b}^k\right]{\bar b}^{M-k-3}+\sum_{k=1}^{M-3}
\left[\Tr{\bar b}^k\right]b{\bar b}^{M-k-2}\right]\ket{0}\nonumber
\eea
\bea
\bra{0}\Tr b^{M-1}{\bar b}^{M-1}\ket{0}&=&\left(N^2-\left[\frac{M-1}{2}\right]^2\right)\bra{0}\Tr b^{M-3}{\bar b}^{M-3}\ket{0}
\nonumber\\
&&+\bra{0}\Tr b^{M-3}\left[-N\sum_{k=odd}^{M-3}
\left[\Tr{\bar b}^k\right]{\bar b}^{M-k-3}+N\sum_{k=odd}^{M-3}
\left[\Tr{\bar b}^k\right]{\bar b}^{M-k-3}\right]\ket{0}\nonumber\\
&&-\bra{0}\Tr b^{M-3}\sum_{k=odd}^{M-3}
\left[\Tr{\bar b}^k\right]\sum_{l=odd}^{M-k-3}\left[\Tr{\bar b}^l\right]{\bar b}^{M-k-l-3}\ket{0}
\nonumber\\
&=&\left(N^2-\left[\frac{M-1}{2}\right]^2\right)\bra{0}
\Tr b^{M-3}{\bar b}^{M-3}\ket{0}
\eea
The last line is reached by noting the cancellation of the sums in
the square brackets and noting that the double sum in the previous line
can be rearranged as
\bea
\left[\Tr{\bar b}^k\right]\sum_{l=odd}^{M-k-3}\left[\Tr{\bar b}^l\right]{\bar b}^{M-k-l-3}\ket{0}&=&\sum_{n=even,2}^{M-3}\left[\sum_{k=odd,1}^{n-1}
\Tr{\bar b}^k\Tr{\bar b}^{n-k}\right]{\bar b}^{M-n-3}\ket{0}
\eea
and the sum within square brackets is zero because the terms cancel in
pairs due to the fact that $\Tr{\bar b}^k,\Tr{\bar b}^{n-k}$
are fermionic operators. So we have proved the recursion relation
\bea
\bra{0}\Tr b^{2n}{\bar b}^{2n}\ket{0}&=&(N^2-n^2)\bra{0}\Tr b^{2n-2}{\bar b}^{2n-2}\ket{0}=N\prod_{k=1}^n(N^2-k^2)
\eea
To apply the variational principle, we should restrict to integer $N$
so the Hamiltonian is truly hermitian, and require $N>n=(M-1)/2$ so
that the norm of the state is positive. With these restrictions we
can then rigorously conclude that $E_G<-2M$ for $M$ odd and all $N>(M-1)/2$.
We gain no information when $N\leq(M-1)/2$. In Appendix B, we explain
an extension of this variational argument to more general
trial states.

The fact that the upper bound on the ground state energy behaves like
$-2M$ is consistent with the ground state having stringy properties
similar to those seen at infinite $N$. The ground energy of the 
latter has the large 
$M$ behavior $-8M/\pi$, consistent with the variational bound. 
It is interesting that the
variational result only applies when $N>(M-1)/2$, which suggests,
but does not prove, that high bit number stringy states may require
very large $N$. We shall find further support for these
conclusions in the study of low $M$ energy eigenstates that follows.
\section{Low $M$ Examples}
The eigenvalues and eigenstates of the Hamiltonian (\ref{bitham}) 
in the 't Hooft
limit $N\to\infty$ are reviewed in Appendix A. Before studying our
model at finite $N$ for specific
values of $M$, it is helpful to note that the case $N=1$ is trivial to
solve. Then there is just a single $a$ and a single $b$ and $H$ reduces to
\bea
H_1&=&2\left[{\bar a}^2a^2+2{\bar a}{\bar b}ba\right]
=2(M^2-M)
\eea 
an explicit function of the bit number operator $M={\bar a}a+{\bar b}b$.
Eigenstates of $M$ are automatically eigenstates of $H$. Indeed for 
fixed $M$ there are only two states:
\bea
\ket{M,b}&=&{\bar a}^M\ket{0},\qquad \ket{M,f}={\bar a}^{M-1}{\bar b}\ket{0},
\eea
the first a boson and the second a fermion. This energy spectrum 
rules out the emergence of string and holography for $N=1$.
There are no large bit number states with excitations that scale as $1/M$.
Moreover, there are no large bit number states with energies that grow
linearly with $M$. On the other hand, at $N=\infty$, both of these
conditions are met. Clearly the energy
spectrum must exhibit dramatic qualitative
changes, as a function of $N$. It is likely that, at fixed $N$ and large enough $M$, there should be states with the qualitative behavior $E\sim M^2$ 
of the $N=1$ case, which would definitely not be string-like. But
this does not by itself forbid the lowest energy states from being 
string-like. 

In the following, we shall mostly restrict our
analysis to the color singlet sector of state space, 
for which the multi-trace
Fock state basis introduced in \cite{thornfock} will be
exploited. This nonorthogonal basis gives a linearly 
independent span of the sector with fixed bit number $M$,
provided that $N$ is sufficiently large.
Setting up the energy eigenvalue problem in
this basis enables a uniform treatment for all
continuous values of $N$, even non integer ones.
In this formulation, the energy eigenvalues of $H$ are contained in the 
eigenvalues of a matrix ${\cal H}(N)$ that is not hermitian.
This will allow us to explore systematically how
the energy spectrum changes as $N$ changes.
For particular values of $N$, this basis set is not
linearly independent--it is over-complete. Indeed, for $N=1$, this
over-completeness is quite dramatic! In these cases the
matrix ${\cal H}$ is too big and determines more eigenvalues than
are physical, some of which can even be complex, because ${\cal H}$
is not hermitian. The over-complete basis inherits a metric
from the original Fock space, which has nonnegative eigenvalues
for integer $N$, but may have some negative eigenvalues (ghosts)
for fractional $N$. At integer $N$ the eigenstates of ${\cal H}$
which do not correspond to true energy eigenstates must have
zero norm under the inherited metric.
In the following we shall confirm this expectation for some particular 
low values of $M$. Interestingly we also find that 
even for noninteger $N$ all eigenstates with complex energy also
have zero norm, although for some values of noninteger $N$ there
are negative norm eigenstates (ghosts).
In any case it is clear that to get a complete
understanding of the spectrum in this general formalism,
it is essential to find not only the energy eigenvalues,
but also the norms of their energy eigenstates. 

\subsection{2 bits}
The one bit case $M=1$ is trivial. The only states are 
${\bar a}_\alpha^\beta\ket{0}$
and ${\bar b}_\alpha^\beta\ket{0}$, and our selected Hamiltonian applied
to them both gives 0. The singlet states at $M=1$ are simply 
$\Tr{\bar a}\ket{0}$ and $\Tr{\bar b}\ket{0}$, with zero
energy for all $N$.

The action of $H$ on the two bit Fock space is not quite so
trivial 
\bea
H{\bar a}^{\beta}_\alpha {\bar a}^{\delta}_\gamma\ket{0}
&=&\frac{2}{N}\left[\delta_\gamma^\beta[{\bar a} {\bar a}-i{\bar b} {\bar b}]_\alpha^\delta
+\delta_\alpha^\delta[{\bar a} {\bar a}
-i{\bar b} {\bar b}]_\gamma^\beta\right]\ket{0}\\
H{\bar b}^{\beta}_\alpha {\bar b}^{\delta}_\gamma\ket{0}
&=&\frac{2}{N}\left[\delta_\gamma^\beta[i{\bar a} {\bar a}
-{\bar b} {\bar b}]_\alpha^\delta
-\delta_\alpha^\delta[i{\bar a} {\bar a}
-{\bar b} {\bar b}]_\gamma^\beta\right]\ket{0}\\
H{\bar a}^{\beta}_\alpha {\bar b}^{\delta}_\gamma\ket{0}
&=&\frac{2}{N}\left[\delta_\gamma^\beta[{\bar a} {\bar b}+{\bar b} {\bar a}]_\alpha^\delta
-\delta_\alpha^\delta[{\bar a} {\bar b}
-{\bar b} {\bar a}]_\gamma^\beta\right]\ket{0}\eea
Tracing on $\beta\gamma$ and on $\delta\alpha$ produces two color singlet
energy eigenstates:
\bea
H\Tr\ {\bar a} {\bar a}\ket{0}&=&4\Tr\ {\bar a} {\bar a}\ket{0},\qquad
H\Tr\ {\bar a} {\bar b}\ket{0}=4\Tr\ {\bar a} {\bar b}\ket{0}\eea
The first is bosonic and the second is fermionic, giving a supersymmetric
degeneracy. The norms of both states are nonzero for all $N$, and
since their energy eigenvalue is independent of $N$, it must
be equal to the known $N=1$ eigenvalue: $2(M^2-M)\to4$ for
$M\to2$.

Tracing over $\alpha\beta$ and $\gamma\delta$ produces
two more singlet states:
\bea
H\Tr{\bar a}\Tr{\bar a}\ket{0}&=&\frac{4}{N}\Tr{\bar a}^2\ket{0}
=\frac{1}{N}H\Tr{\bar a}^2\ket{0}\\
H\Tr{\bar a}\Tr{\bar b}\ket{0}&=&\frac{4}{N}\Tr{\bar a}{\bar b}\ket{0}
=\frac{1}{N}H\Tr{\bar a}{\bar b}\ket{0}\eea
From which we easily construct the zero energy color singlet eigenstates
\bea
\left[\Tr{\bar a}\Tr{\bar a}-\frac{1}{N}\Tr{\bar a}^2\right]\ket{0},
\qquad \left[\Tr{\bar a}\Tr{\bar b}
-\frac{1}{N}\Tr{\bar a}{\bar b}\right]\ket{0},\qquad E=0
\eea
The squared norms of these two states are $2(N^2-1)$ and $(N^2-1)$
respectively. They have zero norm at $N=1$, so they will drop out
of the spectrum at $N=1$ as they must.

Tracing only one pair of indices gives the simplified action
of $H$ on color adjoint states:
\bea
H[{\bar a}^{} {\bar a}^{}]^\delta_\alpha\ket{0}
&=&2[{\bar a} {\bar a}-i{\bar b} {\bar b}]_\alpha^\delta
+\frac{2\delta_\alpha^\delta}{N}\Tr[{\bar a} {\bar a}]\\
H[{\bar b}^{}{\bar b}^{}]^\delta_\alpha\ket{0}
&=&2[i{\bar a} {\bar a}-{\bar b} {\bar b}]_\alpha^\delta
-i\frac{2\delta_\alpha^\delta}{N}\Tr[{\bar a} {\bar a}]\\
H[{\bar a}^{}{\bar b}^{}]^\delta_\alpha\ket{0}
&=&2[{\bar a} {\bar b}+{\bar b} {\bar a}]_\alpha^\delta\\
H[{\bar b}^{}{\bar a}^{}]^\delta_\alpha\ket{0}
&=&2[{\bar b} {\bar a}-{\bar a} {\bar b}]_\alpha^\delta
+\frac{4\delta_\alpha^\delta}{N} \Tr {\bar a} {\bar b}\eea
The eigenstates of this adjoint sector are
\bea
\ket{E,B}_\alpha^\beta&=&
[{\bar a}^{} {\bar a}^{}]^\delta_\alpha\ket{0}-i\frac{E-2}{2}
[{\bar b}^{} {\bar b}^{}]^\delta_\alpha\ket{0}-\frac{\delta_\alpha^\delta}{N}
\Tr[{\bar a}^{} {\bar a}^{}]\ket{0},\qquad E=\pm 2\sqrt{2}\\
\ket{E,F}_\alpha^\beta&=&[{\bar a}^{} {\bar b}^{}]^\delta_\alpha\ket{0}
+\frac{E-2}{2}
[{\bar b}^{} {\bar a}^{}]^\delta_\alpha\ket{0}-\frac{E\delta_\alpha^\delta}{2N}
\Tr[{\bar b}^{} {\bar a}^{}]\ket{0},\qquad E=\pm 2\sqrt{2}\eea
The rest of the eigenstates, orthogonal to all these, are at $E=0$.
The total number of states in the 2 bit Fock space is $2N^4$.
All together there are 2 singlets with $E=4$, $4(N^2-1)$ adjoint states
with $E=\pm 2\sqrt{2}$ states,
and the rest amount to $2(N^2-1)^2$ states with $E=0$.

Notice that the energy eigenvalues in the 2 bit sector are actually
independent of $N$. Only the degeneracies and the eigenstates
depend on $N$. This is very special to this sector, because of its
tiny state space. At $N=\infty$, it is even smaller than the state space of the
first quantized Hamiltonian $h$, due to the cyclic symmetry constraint.
For $M=2$ we have
\bea
h_2&=&
-4\theta_1\frac{d}{ d\theta_{2}}-4\theta_{2}\frac{d}{ d\theta_1}-4
+4\theta_1\frac{d}{ d\theta_1}+4\theta_2\frac{d}{ d\theta_2}
=4(\theta_2-\theta_1)\left(\frac{d}{ d\theta_{2}}-\frac{d}{ d\theta_1}\right)-4
\eea 
The eigenfunctions and eigenvalues are
\bea
1,\ E=-4;\qquad (\theta_2+\theta_1),\ E=-4;\qquad
 (\theta_2-\theta_1),\ E=+4; \qquad\theta_1\theta_2,\ E=+4
\eea
But only the last two satisfy the cyclic symmetry constraint
$\Psi(\theta_1,\theta_2)=-\Psi(\theta_2,\theta_1)$, which accords
exactly with the Fock space analysis.
\subsection{3 bit singlets}
The single trace states in the $M=3$ subspace of states are:
\bea
\Tr{\bar a}^3\ket{0},\quad \Tr{\bar a}{\bar b}^2\ket{0};
\qquad \Tr{\bar a}^2{\bar b}\ket{0},\quad \Tr{\bar b}^3\ket{0}
\eea
and their norms are
\bea
\bra{0}\Tr a^3\Tr{\bar a}^3\ket{0}&=&3\bra{0}\Tr a^2{\bar a}^2\ket{0}
=3N^3+3N\nonumber\\
\bra{0}\Tr b^2a\Tr{\bar a}{\bar b}^2\ket{0}&=&\bra{0}\Tr b^2{\bar b}^2\ket{0}
=N^3-N\nonumber\\
\bra{0}\Tr a^2b\Tr{\bar b}{\bar a}^2\ket{0}&=&\bra{0}\Tr a^2{\bar a}^2\ket{0}
=N^3+N\nonumber\\
\bra{0}\Tr b^3\Tr{\bar b}^3\ket{0}&=&3\bra{0}\Tr b^2{\bar b}^2\ket{0}
=3N^3-3N
\eea
For starters we work out the eigenstates at $N=\infty$, where 
it suffices to find the single trace eigenstates, because
all remaining color singlet energy eigenstates are tensor
products of these at infinite $N$. In addition,
we can work independently in the
Bose and Fermi sectors. The action of $H$ on each basis state
in the Bose sector is given by:
\bea
H\Tr{\bar a}^3\ket{0}&=&\frac{6}{N}\Tr({\bar a}^2 -i{\bar b}^2)a{\bar a}^2
\ket{0}
=6\Tr({\bar a}^2 -i{\bar b}^2){\bar a}\ket{0}
+\frac{6}{ N}\Tr({\bar a}^2)
\Tr{\bar a}\ket{0}\\
H\Tr{\bar a}{\bar b}^2\ket{0}&=&\frac{2}{N}\Tr\left[({\bar a} {\bar b}
+{\bar b} {\bar a})b{\bar b}^2+({\bar a} {\bar b}
-{\bar b} {\bar a})a({\bar b}{\bar a}-{\bar a}{\bar b})
-({\bar b}^2-i{\bar a}^2)b({\bar b}{\bar a}-{\bar a}{\bar b})\right]
\ket{0}\nonumber\\
&=&{2}\Tr\left[
-(3{\bar b}^2-i{\bar a}^2){\bar a}
\right]
\ket{0}+\frac{2}{N}\left[-2\Tr({\bar a} {\bar b})\Tr{\bar b}
+\Tr(-i{\bar a}^2)\Tr{\bar a}\right]
\ket{0}\eea
Taking $N\to\infty$, eigenvalue equation becomes:
\bea
E\Tr(c_1{\bar a}^3+c_2{\bar a}{\bar b}^2)\ket{0}&=&
\Tr[6c_1({\bar a}^2-i{\bar b}^2){\bar a}+c_2(2i{\bar a}^3-6{\bar b}^2{\bar a})]
\ket{0}
\eea
which is easily solved:
\bea
\ket{Eb}_\infty&=&c_1\Tr\left({\bar a}^3+\frac{1}{2i}(E-6){\bar a}{\bar b}^2\right)\ket{0},\qquad E=\pm4\sqrt{3}
\eea
Next we move on to the Fermi sector, for which the action of $H$ gives
\bea
H\Tr{\bar b}^3\ket{0}
&=&-6\Tr({\bar b}^2-i{\bar a}^2){\bar b}\ket{0}
+\frac{6}{ N}\Tr({\bar b}^2-i{\bar a}^2)\Tr{\bar b}\ket{0}\\
H\Tr{\bar a}^2{\bar b}\ket{0}
&=&2\Tr\left[(3{\bar a}^2 -i{\bar b}^2){\bar b}
\right]\ket{0}
+\frac{2}{N}\left[\Tr{\bar a}^2 \Tr{\bar b}
+2\Tr({\bar a}{\bar b})\Tr{\bar a}\right]\ket{0}
\eea
We next find the eigenstates at $N=\infty$:
\bea
\ket{Ef}_\infty&=&c_1\Tr\left({\bar b}^3-\frac{1}{2i}(E+6)\Tr{\bar a}^2{\bar b}
\right)\ket{0},\qquad E=\pm4\sqrt{3}
\eea
The Fermi-Bose degeneracy here and at all levels is due to the
existence of the supercharge $Q=\Tr[e^{i\pi/4}{\bar a}b+ e^{-i\pi/4}{\bar b}a]$
which commutes with the Hamiltonian. $[Q,H]=0$.

\begin{figure}[ht]
\begin{center}
\includegraphics[width=6.5in]{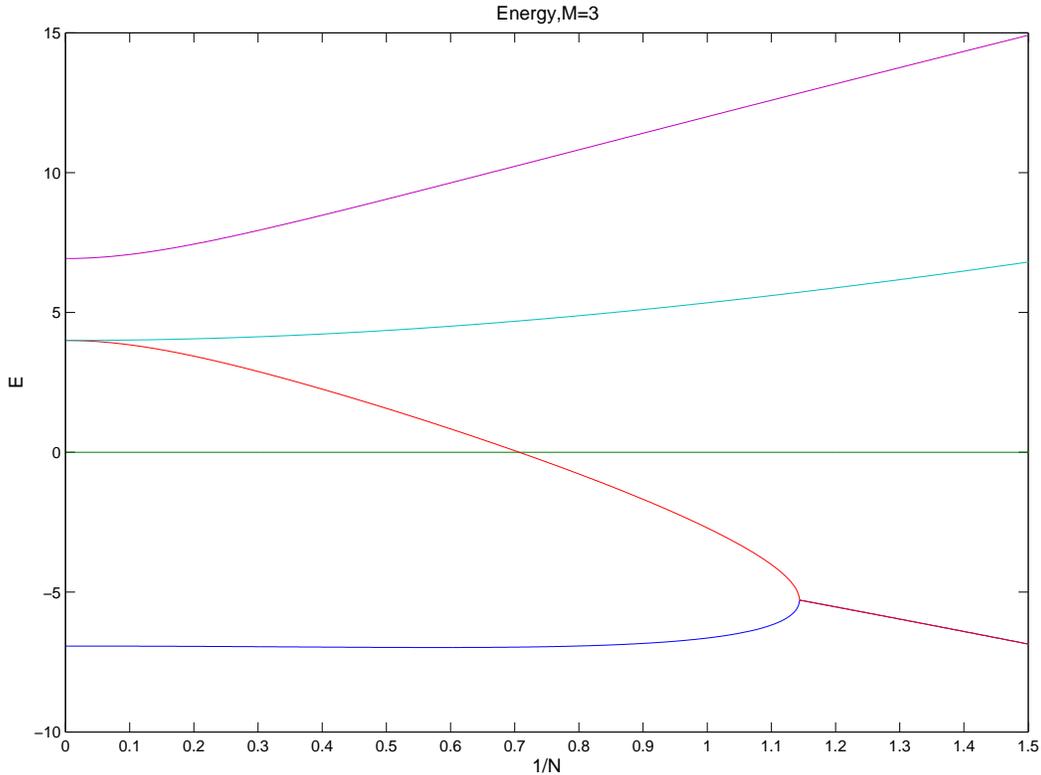}
\caption{The energy eigenvalues of the 3 bit system as a function
of $1/N$. The two energy curves which merge into a single curve,
for $N$ a bit less than 1, actually become complex conjugate pairs 
upon merging. As seen in the next figure, those states have zero norm
when they become complex.}
\label{energy3}
\end{center}
\end{figure}

To go beyond the $N=\infty$ limit, we require the action of $H$ on the
other states in the 3 bit Fock space. Staying within the color
singlet sector, we label the states as:
\bea
\ket{1}&=&\Tr{\bar a}^3\ket{0}
\\
\ket{2}&=&\Tr{\bar a}^2\Tr{\bar a}\ket{0}
\\
\ket{3}&=&\Tr{\bar a}\Tr{\bar a}\Tr{\bar a}\ket{0}
\\
\ket{4}&=&\Tr{\bar a}{\bar b}^2\ket{0}
\\
\ket{5}&=&\Tr{\bar a}{\bar b}\Tr{\bar b}\ket{0}
\eea
Then the result of applying the Hamiltonian on each state can be expressed as
\\
\bea
H\ket{i}&=&\sum_{j}\ket{j}{\cal H}_{ji}
\eea
where the matrix $\cal H$ is 
\bea
{\cal H}&=&\begin{pmatrix}6&\frac{8}{N}&0&2i&0\\ \frac{6}{N}&4&\frac{12}{N}&
-\frac{2i}{N}&0\\ 0&0&0&0&0
\\-6i&-\frac{8i}{N}&0&-6&-\frac{4}{N}\\0&0&0&-\frac{4}{N}&4
\end{pmatrix}\neq {\cal H}^\dagger
\eea
We show the eigenvalues of ${\cal H}$ plotted against $1/N$ 
in Fig.~\ref{energy3}.
The eigenvalues of ${\cal H}$ and their eigenstates will be in
one to one correspondence with the eigenvalues and eigenstates of $H$
in the three bit sector, provided the basis set $\ket{i}$ is linearly 
independent. If there are linear dependences among the $\ket{i}$,
the eigenstates of $H$ will correspond to a proper
subset of those of ${\cal H}$.
To assess linear independence we also need to
calculate the metric $G_{ij}=\VEV{i|j}$ which is given by
\bea
G&=&\begin{pmatrix}3+\frac{3}{N^2}&\frac{6}{N}&\frac{6}{N^2}&0&0\\ 
\frac{6}{N}&2+\frac{4}{N^2}&\frac{6}{N}&
0&0\\ \frac{6}{N^2}&\frac{6}{N}&6&0&0
\\0&0&0&1-\frac{1}{N^2}&0\\0&0&0&0&1-\frac{1}{N^2}
\end{pmatrix}=G^\dagger
\eea
The eigenvalues of $G$ will be real and nonnegative for integer values
of $N$, because the state space acted on by $H$ is positive
definite. But $G$ will have a zero eigenvalue corresponding to
each linear dependence in the basis. When $N$ is not an integer,
there can be both zero and negative eigenvalues of $G$, because
there is no physical state space for such values of $N$. If we
write, for an eigenvector of ${\cal H}$, 
$\ket{E}=\sum_i\ket{i}v^i$, its norm can be expressed in terms of the
matrix $G$:
\bea
\VEV{E|E}&=&\sum_{ij}v^{i*}\VEV{i|j}v^j=v^\dagger Gv
\eea 
\begin{figure}[ht]
\begin{center}
\includegraphics[width=6.5in]{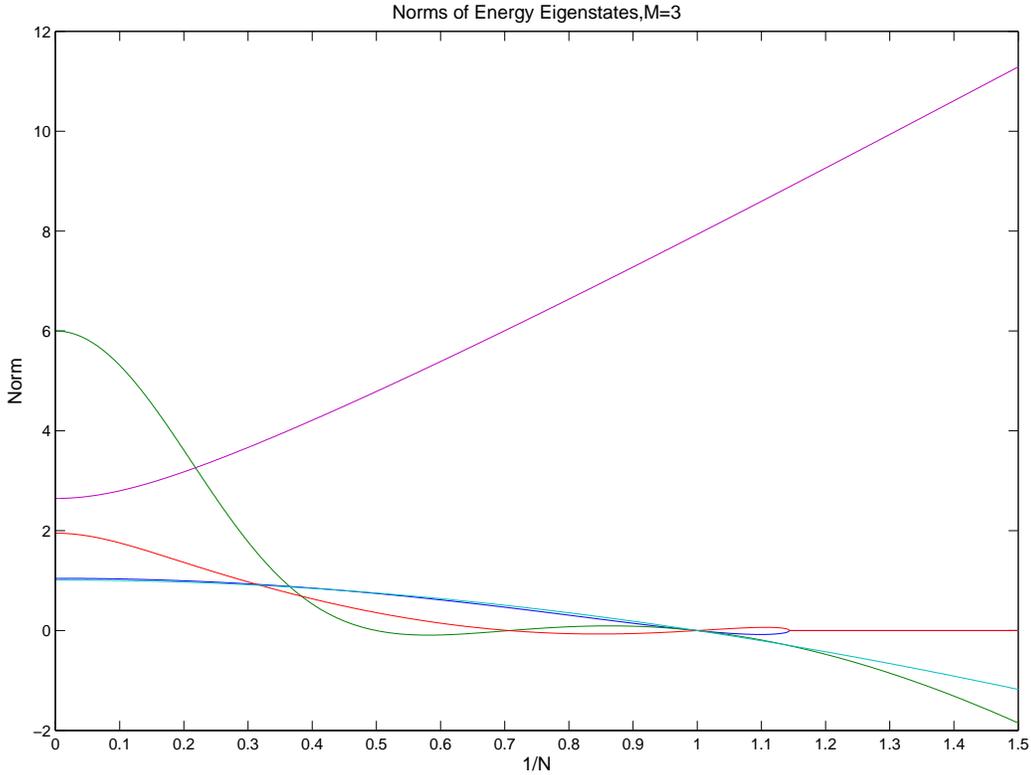}
\caption{The norms of the energy eigenstates of the 3 bit
system as a function of
$1/N$. The color of the norm curve is the same as the corresponding
energy curve. Note that all states but the one with highest energy
have zero norm at $N=1$. This agrees with our exact treatment of
that case. Note the presence of negative norms (ghosts) when $N<2$.}
\label{norms3}
\end{center}
\end{figure}
For integer $N$ this eigenvalue will be in the spectrum of $H$ provided
$\VEV{E|E}\neq0$. In Fig.~\ref{norms3}, we plot this norm for
each of the eigenvalues shown in Fig.~\ref{energy3}. 
Whenever the norm is zero the eigenvalue $E$ is not in the spectrum of $H$,
and in that case it need not even be real, because ${\cal H}$ is
not hermitian. However it is not hard to see that ${\cal H}^*$
is similar to ${\cal H}$, so that complex eigenvalues always 
occur in complex conjugate pairs. This is what is happening when two
eigenvalue curves merge as seen in Fig.~\ref{energy3}. To the right
of the merger we plot only the real part of $E$, and the two eigenvalues
have equal and opposite imaginary parts.

At $N=\infty$, the energy eigenvalues  are 
$E=\pm8\sin(\pi/3)=\pm4\sqrt{3}; E=4; E=4;$ and $E=0$.
The first two eigenvalues come from the single trace states,
The next two come from the double trace states, which agree 
with the 2-bit results, and the triple trace state 
$\Tr{\bar a}\Tr{\bar a}\Tr{\bar a}\ket{0}$ has zero
energy.

Notice that at $N=1$ the five curves for the energies,
Fig.\ref{energy3}, approach distinct values.
On the other hand, Fig.~\ref{norms3} shows that the norms of the lowest
four of them are $0$ at $N=1$. These states then disappear from the
spectrum at $N=1$: only the state evolving from $Tr{\bar a}^3\ket{0}$ 
at $N=\infty$ survives at this value of $N$.
Also at $N=1$, $H_1=2(M^2-M)=12$ which agrees with the top
curve of Fig.\ref{energy3}.

At $N=2$ the norm of the  state, which evolves from 
$\Tr{\bar a}\Tr{\bar a}\Tr{\bar a}\ket{0}$, goes to zero, and between $N=2$ 
and $N=1$ its norm goes negative for a while. The presence
of such a ghost state, which can happen only for
noninteger $N$, signals a violation of unitarity. Even though At
$M=3$ this disappearing state is a rather trivial 
one with zero energy eigenvalue, we have found in higher $M$ cases 
that the norms of nontrivial states go to zero at integer $N$. This
suggests a pattern that as $M$ goes up, the states start to disappear at 
higher integer $N$.

When $M= 3$ no complex conjugate pair appears in the range $N\geq1$. 
But as Fig.~\ref{energy3} shows, two curves do merge in the region $N<1$.
At the merge point the eigenvalues become complex conjugate pairs. 
At higher $M$, such complex energies occur at higher $N$.  
In our detailed studies at $M=4$ and $M=5$ we find that every eigenstate
with complex energy has zero norm, indicating that it disappears from 
the spectrum, whether $N$ is an integer or not. 
This has to happen at integer
$N$ where $H$ is manifestly hermitian, but not 
necessarily at unphysical fractional $N$. The fact
that it does indicates that unitarity may be possible for
a range of continuous $N$. In the string
interpretation of these models $1/N$ acts as the
string coupling constant, which is not {\it a priori} quantized.
So in this context continuous $N$ might have some physical meaning.
Even so our studies do find ranges of noninteger $N$ with negative
norm eigenstates (ghosts), so unitarity is by no 
means assured at fractional $N$.

We can see that the $N$ dependence of the ground state energy in the
three bit sector is very flat from $N=\infty$ to $N=2$, 
and only goes up a little bit at $N=1$. This is also true for the $M=5$ case,
suggesting that the string-like properties of this
state may persist, to some extent, as $N$ decreases from $\infty$.
For $M=4$ (and all even $M$) this flat ground state is missing 
because it doesn't satisfy the cyclic symmetry constraint. 
The detailed analysis of the singlet spectrum for $M=4$, $M=5$ and
higher will appear in a separate article. 

\subsection{3 bit adjoints} 
We briefly consider the $N$ dependence of the energy spectrum
in the color adjoint sector. In general an adjoint state can
be characterized by a monomial of creation operators ${\bar A}_\alpha^\beta$
carrying free color indices. Then the adjoint state would be
\bea
{\bar A}_\alpha^\beta\ket{0}-\frac{\delta_\alpha^\beta}{N}\Tr {\bar A}\ket{0}.
\eea
A convenient metric for these states can
be taken to be
\bea
\bra{0}\Tr B{\bar A}\ket{0}-\frac{1}{N}\bra{0}\Tr B\Tr{\bar A}\ket{0}.
\eea
Specializing to 3 bits ($M=3$) we can then form the following monomials
\bea
{\bar A}_1&=&[{\bar a}{\bar a}{\bar a}],\quad {\bar A}_2=[{\bar b}{\bar b}{\bar a}],\quad {\bar A}_3=[{\bar a}{\bar b}{\bar b}],\quad {\bar A}_4=[{\bar b}{\bar a}{\bar b}]\\
{\bar A}_5&=&[{{\bar a}}{{\bar a}}]Tr[{\bar a}],\quad {\bar A}_6=[{\bar a}]Tr[{\bar a}{\bar a}],\quad {\bar A}_7=[{\bar b}{\bar b}]Tr[{\bar a}]\\ 
{\bar A}_8&=&[{\bar a}{\bar b}]Tr[{\bar b}],\quad {\bar A}_9=[{\bar b}{\bar a}]Tr[{\bar b}],\quad {\bar A}_{10}=[{\bar b}]Tr[{\bar a}{\bar b}]\eea
where square brackets surround the monomial carrying the free
color indices.
Then the analogue of the multi-trace basis is just
\bea
\ket{i}_\alpha^{~\beta}&=&({{\bar A}}_i)_\alpha^{~\beta}\ket{0}.
\eea
The Hamiltonian matrix of the adjoint states in the 3 bit
sector is defined by
\bea
H\ket{i}&=&\sum_j\ket{j}{\cal H}^A_{ji}
\eea
with
\bea
{\cal H}^A&=&\begin{pmatrix}4&2i&2i&0&\frac{8}{N}&\frac{8}{N}&0&0&0&0\\
 -2i&-4&0&2&-\frac{4i}{N}&-\frac{4i}{N}&0&0&0&-\frac{4}{N}\\
 -2i&0&0&2&-\frac{4i}{N}&-\frac{4i}{N}&\frac{4}{N}&-\frac{4}{N}&0&\frac{4}{N}\\
 0&2&2&0&0&0&\frac{4}{N}&0&\frac{4}{N}&-\frac{4}{N}\\
 \frac{2}{N}&0&0&\frac{2i}{N}&2&0&2i&0&0&0\\
 \frac{4}{N}&-\frac{2i}{N}&-\frac{2i}{N}&0&0&4&0&0&0&0\\
 -\frac{2i}{N}&0&0&-\frac{2}{N}&-2i&0&-2&0&0&0\\
 0&\frac{2}{N}&-\frac{2}{N}&0&0&0&0&2&2&0\\
 0&-\frac{2}{N}&-\frac{2}{N}&0&0&0&0&2&-2&0\\
 0&\frac{4}{N}&0&-\frac{4}{N}&0&0&0&0&0&4\\
\end{pmatrix}
\eea

\begin{figure}[ht]
\begin{center}
\includegraphics[width=6in]{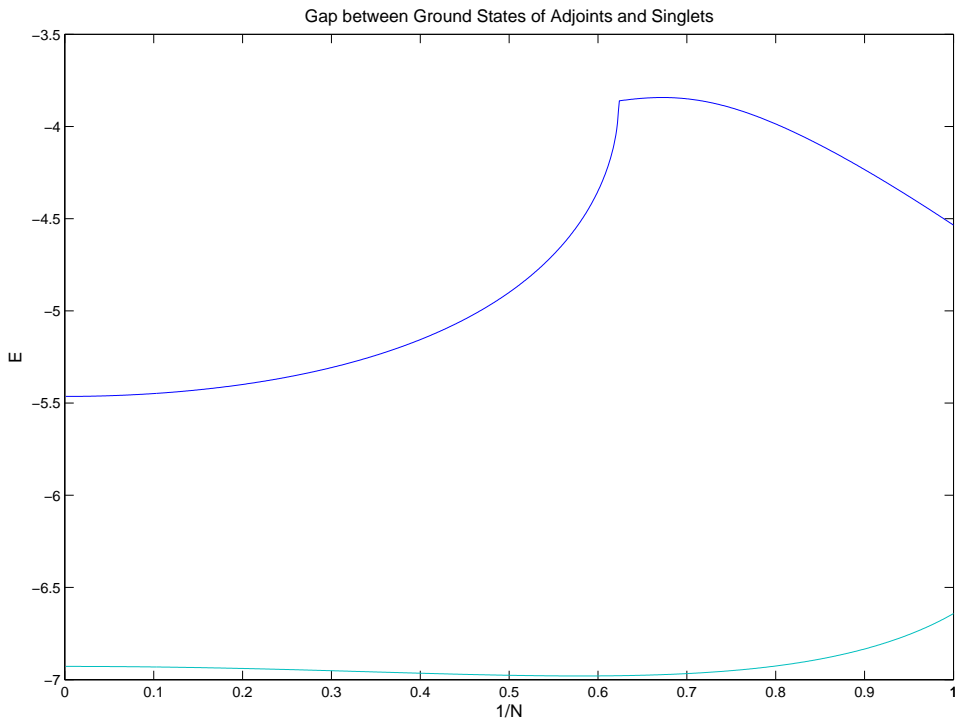}
\caption{The lowest energies of the color adjoint (top curve) and 
color singlet (bottom curve)
states in the 3 bit sector, showing a gap that persists in the
entire range $1\leq N<\infty$. The cusp is really the point where
two real energy curves merge and the energies to the right of the cusp have
equal and opposite imaginary parts.}
\label{gap3}
\end{center}
\end{figure}

The details of the adjoint spectrum will be described in a separate
publication. Here we just show the gap between adjoint and
singlet states by plotting in Fig.~\ref{gap3}   
the lowest energy
eigenvalues in each of these color representations.
It is interesting that at least at $M=3$ there is no tendency
for the gap to close as $N$ decreases from infinity.
The cusp seen in the lowest energy adjoint curve is the point
at which the next highest adjoint energy merges with the
lowest adjoint energy and thereafter both energies become
complex with equal and opposite imaginary parts. To the right
of the cusp the lowest real adjoint energy is yet higher.

\section{Extensions and Concluding Remarks}
In this article we have studied the simplest superstring bit model,
which underlies the $s=1$, $d=0$ superstring. As mentioned in the
introduction, the extension to a superstring bit model underlying the
$s>1$, $d=0$ superstring is straightforward. One simply enlarges the
set of bit creation operators to 128 bosonic and 128 fermionic ones
\bea
a_{\alpha}^{~\beta}, b_{\alpha}^{~\beta} \to 
(\phi_{[a_1,\cdots,a_n]})_{\alpha}^{~\beta}, n=0,\ldots,s 
\eea
where each $a_k$ is a spinor index running over $s$ values.
The appropriate Hamiltonian whose 't Hooft limit gives the
$s=8$, $d=0$ superstring can be obtained from the one constructed
in \cite{bergmantsubit} by dropping all of the transverse coordinate
dependence:
\bea
H&=&\frac{1}{N}\sum_{n=0}^s\frac{s-2n}{n!}
\Tr{\bar\phi}_{a_1\cdots a_n}\rho{\phi}_{a_1\cdots a_n}\nonumber\\
&&+\frac{1}{N}\sum_{n=0}^{s-1}\frac{1}{n!}
\Tr{\bar\phi}_{a_1\cdots a_n}\eta_b{\phi}_{ba_1\cdots a_n}
+\frac{1}{N}\sum_{n=0}^{s-1}\frac{1}{n!}
\Tr{\bar\phi}_{ba_1\cdots a_n}{\bar\eta}_b{\phi}_{a_1\cdots a_n}
\eea
where we have defined:
\bea
\rho&=&\sum_{k=0}^s\frac{1}{k!}{\bar\phi}_{b_1\cdots b_k}
{\phi}_{b_1\cdots b_k}\\
\eta_b&=&\sum_{k=0}^{s-1}\frac{(-)^k}{k!}{\bar\phi}_{bb_1\cdots b_k}
{\phi}_{b_1\cdots b_k}+i\sum_{k=0}^{s-1}\frac{(-)^k}{k!}{\bar\phi}_{b_1\cdots b_k}{\phi}_{bb_1\cdots b_k},\qquad {\bar\eta}_b=-i\eta_b
\eea
Of course when this expression is specialized to $s=1$, it is just the
Hamiltonian we have analyzed in this article. When $s=8$ its
large $N$ limit just describes the spinor sector of the critical superstring
in light cone gauge.
For any $s>0$ the model can be analyzed in a manner exactly parallel to the
analysis in this article. We defer this generalization to 
a future publication. Only one spatial coordinate, $x^-$,
will be holographically generated in the large $N$ limit, and
the excitations of the superstring will be described by $s$ lightcone
worldsheet Grassmann coordinates $\theta_a$.

The task of constructing a superstring bit model that underlies
superstring theory with $d>0$ calls for some interesting choices.
The model constructed in \cite{bergmantsubit} to describe the
$s=8$, $d=8$ superstring addressed this problem
by simply promoting the quantum mechanical variables $\phi\to\phi({\bfs x})$
to fields on the $d$ dimensional transverse space ${\bfs x}$. The Hamiltonian
for that model included two body terms, quartic in the $\phi$'s
and involving a potential $V({\bfs x}-{\bfs y})$. To exactly produce
a harmonic superchain, in the 't Hooft limit, required a harmonic
oscillator potential $V=T_0({\bfs x}-{\bfs y})^2$. We regarded this 
long range potential as unsatisfactory. After all the transverse space
for the bits was identified with the transverse space after the
holographic emergence of $x^-$, and we felt locality in the dynamics
in the emergent $d+2$ dimensional space-time would be unlikely unless the 
potential between bits was short range. In subsequent papers
\cite{bergmantsugal} 
we struggled mightily, without complete success, to build 
satisfactory superstring bit models with a short range potential.

But in retrospect this effort seems philosophically misguided.
The principal motivation for string bit models is to replace
quantum field theory, with an infinite number of degrees of
freedom, with an underlying theory with a finite number of
degrees of freedom. Letting the string bits move in transverse space
reinserts an infinite number of degrees of freedom in the
underlying theory. It is philosophically more coherent to
seek a model in which all dimensions of space emerge holographically.
We can easily see how this can happen. It has been known for
a long time that the Heisenberg chain of spins. 
\bea
H_{hei}=-\frac{1}{2}\sum_k\left(\sigma_k^1\sigma^1_{k+1}+\sigma_k^2\sigma^2_{k+1}
+\Delta\sigma_k^3\sigma^3_{k+1}\right)
\eea
describes in the continuum limit a spatial coordinate 
compactified on a circle of
radius $R=2\pi/\sqrt{2T_0(\pi-\mu)}$ where $\Delta=-\cos\mu$ \cite{gilesmt}. 

It is easy to incorporate this idea in string bit models.
Append a two valued ``flavor index'' for  each
transverse dimension: $\phi_{[a_1\cdots a_n]}\to 
\phi_{[a_1\cdots a_n]}^{f_1\cdots f_d}$, with $f_i=1,2$.
and design the string bit Hamiltonian to produce the Heisenberg
Hamiltonian on the long chains that naturally
arise in the limit $N\to\infty$. Such a string bit model has
$2^s2^dN^2$ ($=(256N)^2$ for the superstring bit model) degrees of freedom.
Pursuit of these ideas is an exciting project for future investigation.
\vskip14pt
\noindent\underline{Acknowledgments}: We would like to thank 
Philip Mannheim and Lars Brink for helpful discussions.
This research was supported in part by the Department
of Energy under Grant No. DE-FG02-97ER-41029.

\appendix
\section{Diagonalizing $H$ in the Large $N$ limit}
\subsection{Color Singlets}
We have shown that the action of $H$ on single trace singlet states
can be described in the large $N$ limit in terms of the ``first-quantized''
Hamiltonian $h$. To find the eigenvalues of $h$,
it is convenient to introduce Fourier transforms
\bea
\alpha_n&=&\frac{1}{\sqrt{M}}\sum_{k=1}^M\theta_k e^{2\pi ikn/M},\qquad
\beta_n=\frac{1}{\sqrt{M}}\sum_{k=1}^M\frac{d}{d\theta_k} e^{2\pi ikn/M}\\
\theta_k&=&\frac{1}{\sqrt{M}}\sum_{n=0}^{M-1}\alpha_n e^{-2\pi ikn/M},\qquad
\frac{d}{d\theta_k}=\frac{1}{\sqrt{M}}\sum_{n=0}^{M-1}\beta_n e^{-2\pi ikn/M}\\
\{\alpha_n,\beta_m\}&=&\delta_{m+n,M}
\eea
The we can express $h$ in terms of these
\bea
h&=&-2M+2\sum_{n=1}^{M-1}\left[(\alpha_n\alpha_{M-n}+\beta_n\beta_{M-n})
\sin\frac{2\pi n}{M}+(\alpha_n\beta_{M-n}+\alpha_{M-n}\beta_{n})
\left(1-\cos\frac{2\pi n}{M}\right)\right]\nonumber\\
&=&-2M+4\sum_{n=1}^{M-1}\sin\frac{n\pi}{M}\left[(\alpha_n\alpha_{M-n}+\beta_n\beta_{M-n})
\cos\frac{\pi n}{M}+(\alpha_n\beta_{M-n}+\alpha_{M-n}\beta_{n})
\sin\frac{\pi n}{M}\right]
\eea
For $n<M/2$, we now diagonalize the operator in square brackets, which
we call $[\phantom{0}]_n$. The functions $\alpha_n$ and $\alpha_{M-n}$
are eigenfunctions of $[\phantom{0}]_n$ with eigenvalue $\sin(n\pi/M)$.
The remaining two eigenfunctions are of the form
$a+b\alpha_n\alpha_{M-n}$:
\bea
[\phantom{0}]_n(a+b\alpha_n\alpha_{M-n})&=&a\alpha_n\alpha_{M-n}\cos\frac{\pi n}{M}
+b\left[\cos\frac{\pi n}{M}
+2\alpha_n\alpha_{M-n}\sin\frac{\pi n}{M}\right]
=\epsilon_n(a+b\alpha_n\alpha_{M-n})\nonumber\\
b&=&\frac{a\epsilon_n}{\cos(n\pi/m)},\qquad \cos^2\frac{\pi n}{M}
+2\epsilon_n\sin\frac{\pi n}{M}=\epsilon_n^2\nonumber\\
\epsilon_n&=&\sin\frac{\pi n}{M}\pm1
\eea 
When $M$ is even, there is a term with $n=M/2$ for which the operator
in square brackets is simply
\bea
[\phantom{0}]_{M/2}&=&2\alpha_{M/2}\beta_{M/2}
\eea
whose eigenfunctions are $1$ with value $0=\sin(\pi/2)-1$ and $\alpha_{M/2}$ 
with  value $2=\sin(\pi/2)+1$. If we take the common term $\sin(n\pi/M)$
for each eigenvalue, we see that it contributes
\bea
4\sum_{n=1}^{M-1}\sin^2(n\pi/M)=2M
\eea
to the eigenvalue of $h$. This simply cancels the $-2M$ term in $h$.
Thus we may write the general
eigenvalue of $h$ as
\bea
E(\{\eta_n\})&=&8\sum_{n<M/2}\eta_n\sin\frac{n\pi}{ M}+4\eta_{M/2}\\
\eta_n&=&+1,0,0,-1,\quad {\rm for}\quad n\neq\frac{M}{2},\qquad
\eta_{M/2}=\pm1
\eea
Clearly the smallest and largest eigenvalues are
\bea
E_{\rm max}&=&4\sum_{n=1}^{M-1}\sin\frac{n\pi}{ M}
=4\cot\frac{\pi}{2M},\qquad E_{\rm min}=-E_{\rm max}
\eea
In the limit of many bits, $M\to\infty$ we have the behavior
\bea
E_{\rm max}&=&\frac{8M}{\pi}-\frac{2\pi}{3M}+{\mathcal O}(M^{-3})\\
E_{\rm min}&=&-\frac{8M}{\pi}+\frac{2\pi}{3M}+{\mathcal O}(M^{-3})
\eea
The term linear in $M$ can be cancelled against a counter term in the
Hamiltonian of the form $(8/\pi)\Tr[{\bar a}a+{\bar b}b]$. 
Then interpreting $M$ as a discretized $P^+=Mm$ identifying
$P^-=ET_0/m$ and we see that the spectrum is relativistic in the
limit $M\to\infty$.

Finally we consider the cyclic symmetry requirements. Under a single step cycle
$\alpha_n\to e^{2in\pi/M}\alpha_n$ so the product $\alpha_n\alpha_{M-n}$
is invariant. When $M$ is odd, the states are supposed to be invariant under
such a cycle. This is realized when the values of $\eta_n$ are restricted to $\pm1$, or, more generally when a number of the $\eta_k=0$ 
for which $\sum_{k,\eta_k=0}k=M$.
When $M$ is even the states must change sign under a single step cycle.
This means that the $M/2$ wave function must be $\alpha_{M/2}$,
so $\eta_{M/2}=+1$ and all the other $\eta_n=\pm1$, or for those
with $\eta_k=0$, $\sum_k k=M$. In particular
the eigenvalue $E_{\rm min}$ would be excluded for $M$ even but
allowed for $M$ odd. For very large $M$, so the chains behave as continuous
string, the gap between the even and odd $M$ sectors becomes large
compared to the excitation energies of the odd $M$ sector. This means
that only the odd $M$ closed chains will participate in the continuum
physics. This implies a multiplicatively conserved parity symmetry
that forbids an odd number of chains transforming into an even number
of chains. In particular a single chain could not decay into two chains. 

On the other hand $E_{\rm max}$ is allowed
for all $M$. If we want a system for which the ground energy 
at infinite $N$ is allowed for 
all $M$ we could choose its Hamiltonian to be $-h$ rather than $h$.
But this would make the coefficient of the $\Tr{\bar a}^2a^2$
term negative, which threatens a dangerous instability since this
implies attractive interactions between bosons. In \cite{thornsakh}
we noted that adding a term $\Tr{\bar a}a{\bar a}a$ with a positive
coefficient could stabilize the theory at the expense of complicating the
large $N$ analysis. In $h$ the attractive
interactions are between fermions which are tamed by the exclusion principle,
without such complications.
\subsection{Color Adjoints}
For completeness, we also consider the color adjoint states in
the large $N$ limit. In this case we apply $H$ to states of the
form
\bea
\ket{\theta_1,\cdots\theta_M}_\alpha^\beta
&=&[\psi(\theta_1)\cdots\psi(\theta_M)]_\alpha^\beta\ket{0}
\eea
and determine in the large $N$ limit the ``first quantized'' $h_A$
such that
\bea
H\int d^M\theta\ket{\theta_1,\cdots\theta_M}_\alpha^\beta
\Psi(\theta_1,\cdots\theta_M)=\int d^M\theta\ket{\theta_1,
\cdots\theta_M}_\alpha^\beta
h_A\Psi(\theta_1,\cdots\theta_M)+{\mathcal O}(N^{-1}).
\eea
Following similar steps as for the singlets, we find
\bea
h_A&=&\sum_{k=1}^{M-1}\left[-2i\theta_k\theta_{k+1}
-2i\frac{d}{d\theta_k}\frac{d}{d\theta_{k+1}}
-2\theta_k\frac{d}{d\theta_{k+1}}-2\theta_{k+1}\frac{d}{d\theta_k}-2+4\theta_k\frac{d}{d\theta_k}\right].
\eea
We note that, in comparison to $h$, the only change is the deletion
of the term $k=M$. This breaks the closed chain of bits to form an open
chain of bits. To diagonalize $h_A$, we replace periodic 
boundary conditions with
the ones implied by the absence of this term. The net effect of this
change of boundary conditions is to change the mode energies from
$4\sin(n\pi/M)$ to $4\sin(n\pi/2M)$. Correspondingly, the ground state
energy in the adjoint sector at $N=\infty$ is
\bea
E_A=-4\sum_{n=1}^{M-1}\sin\frac{n\pi}{2M}=-2\cot\frac{\pi}{4M}+2
= -\frac{8M}{\pi}+2+\frac{\pi}{6M}+{\mathcal O}(M^{-3}).
\eea
The second term on the right shows that the large $N$ dynamics 
leads to color confinement The energy gap between the
adjoint and singlet sectors at $N=\infty$ is
\bea
E_A-E_G&=& 2-2\tan\frac{\pi}{4M}= 2-\frac{\pi}{2M}+{\mathcal O}(M^{-3}).
\eea
\begin{figure}[ht]
\begin{center}
\includegraphics[width=6in]{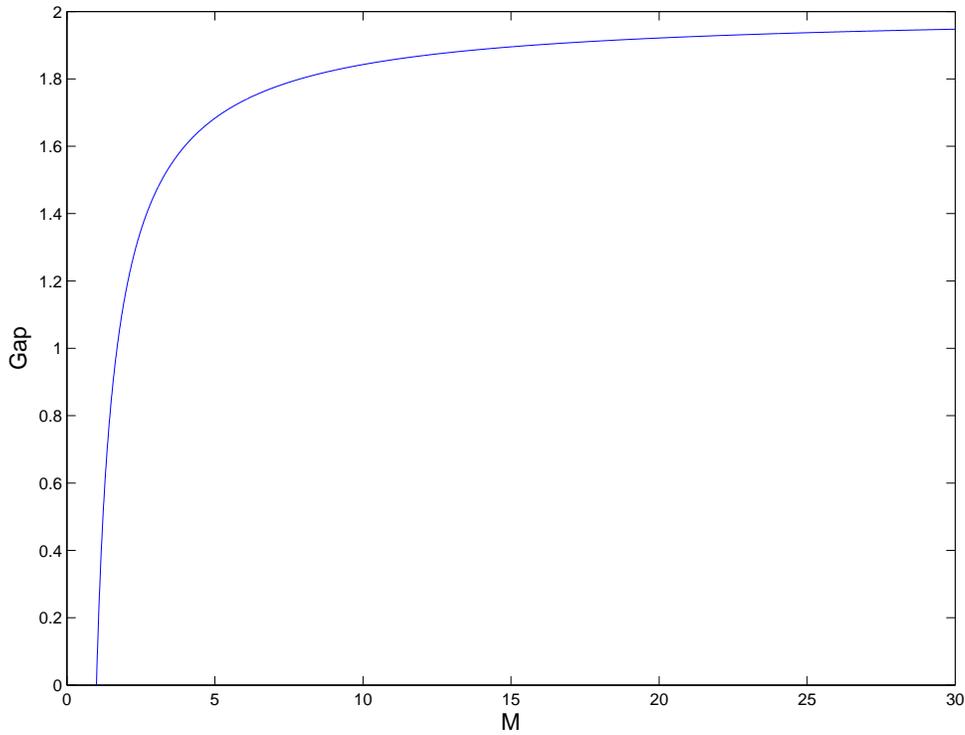}
\caption{The energy gap between color adjoint and color singlet
sectors at $N=\infty$.}
\label{gapinfty}
\end{center}
\end{figure}
In Fig.~\ref{gapinfty} we plot this energy gap as a function of $M$.
The gap remains finite as $M\to\infty$. But
the $1/M$ terms set the scale of energy of the continuum string
excitations. Thus the energy gap between adjoint and singlet sectors
becomes infinitely large in comparison to this string
energy scale when $M\to\infty$. Note that ``perfect'' confinement depends on
$M\to\infty$. If $M$ is simply extremely large rather than
$\infty$, the mass gap is of order $M$ times the scale set by
the $1/M$ excitations.
\section{Truncation to Single Trace States}
We mention here an extension of the variational method described
in Section 3. One can take a trial state to be any linear combination
of single trace states, and vary  the energy function
with respect to the coefficients in this linear combination.
Requiring that the energy function is stationary then implies
that the coefficient functions satisfy the eigenvalue equation
\bea
\sum_j \bra{k}H\ket{j}c_j&=&E\sum_j\VEV{k|j} c_j
\eea
where $j$ is summed over the selected states and $E$ is the energy function.

\begin{figure}[ht]
\begin{center}
\includegraphics[width=4in]{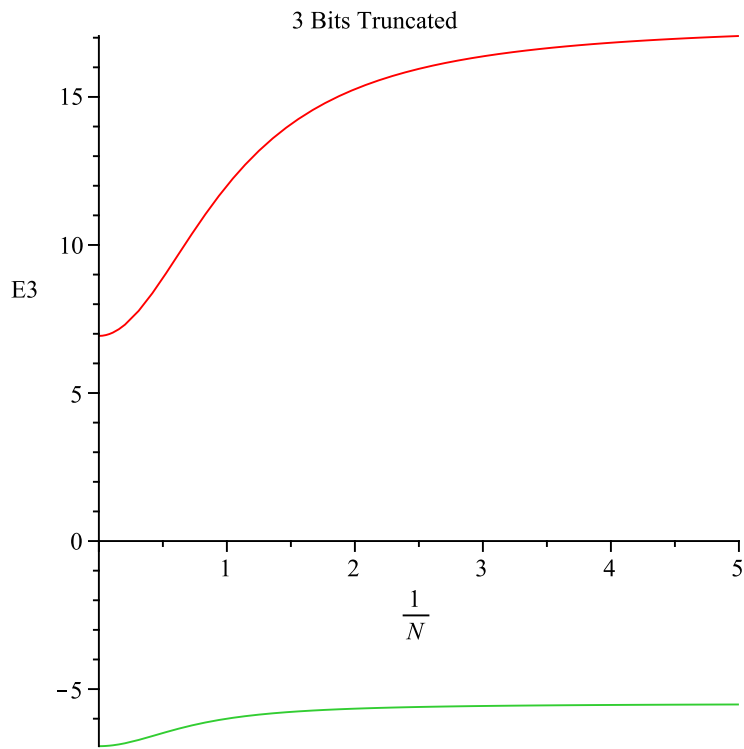}
\caption{Variational estimate of two energy eigenvalues of the 3 bit
system.}
\label{3trunc}
\end{center}
\end{figure}

As a simple example consider the 3 bit case in the boson 
sector. Then there are only two single trace states 
\bea
\ket{1}=\frac{1}{\sqrt{3N(N^2+1)}}\Tr{\bar a}^3\ket{0},\qquad
\ket{2}=\frac{1}{\sqrt{N(N^2-1)}}\Tr{\bar a}{\bar b}^2\ket{0}
\eea
which we normalized and happen to be orthogonal. Thus the factor
$\VEV{k|j}=\delta_{kj}$ and $E$ is then just one of the eigenvalues
of the $2\times2$ matrix
\bea
H&=&\begin{pmatrix}6\frac{N^2+3}{N^2+1}&-2i\sqrt{3}\sqrt\frac{N^2-1}{N^2+1}\\
2i\sqrt{3}\sqrt\frac{N^2-1}{N^2+1}&-6\end{pmatrix}
\eea
The eigenvalues are the roots of aa quadratic polynomial:
\bea
E&=&\frac{6}{ N^2+1}\pm\sqrt{48+\frac{36}{(N^2+1)^2}+\frac{48}{N^2+1}}
\eea
We plot these eigenvalues as a function of $1/N$ in Fig.~\ref{3trunc}.
Of course the curves go to the exact eigenvalues at $N=\infty$. 
It is evident that the lowest of these estimates varies more
steeply than the exact eigenvalue as $N$ decreases from $\infty$.


\end{document}